\documentclass[12pt,preprint,aps,showpacs,showkeys,floatfix]{revtex4}
\usepackage{bm}
\usepackage{amssymb}
\usepackage{amsmath}
\usepackage{amsfonts}
\usepackage{graphicx}
\usepackage{graphics}
\usepackage{float}
\usepackage{epsf}
\usepackage{epsfig}
\usepackage{amssymb}
\usepackage{graphicx}
\usepackage{epsf}

\begin{document}

\newcommand{\be}{\begin{equation}}
\newcommand{\ee}{\end{equation}}
\newcommand{\bea}{\begin{eqnarray}}
\newcommand{\eea}{\end{eqnarray}}
\newcommand{\rr}{{\bf r}}
\newcommand{\rt}{\rr^\perp}
\newcommand{\kk}{{\bf k}}
\newcommand{\kkt}{\kk^\perp}
\newcommand{\p}{{\bf p}}
\newcommand{\q}{{\bf q}}
\newcommand{\qt}{\q^\perp}
\newcommand{\X}{({\bf x})}
\newcommand{\Y}{({\bf y})}
\newcommand{\x}{{\bf x}}
\newcommand{\ff}{{\bf f}}
\newcommand{\uu}{{\bf u}}
\newcommand{\xx}{{\bf x}}
\newcommand{\EE}{{\bf E}}
\newcommand{\VV}{{\bf V}}
\newcommand{\y}{{\bf y}}
\newcommand{\U}{{\bf u}}
\newcommand{\w}{{\bf \omega}}
\newcommand{\D}{{\bf \nabla}}
\newcommand{\W}{\omega_\kk}
\newcommand{\za}{\alpha}
\newcommand{\zb}{\beta}
\newcommand{\zd}{\delta}
\newcommand{\zg}{\gamma}
\newcommand{\zl}{\lambda}
\newcommand{\zs}{\sigma}
\newcommand{\zt}{\tau}
\newcommand{\zE}{I\hskip-3.7pt E}
\newcommand{\zN}{I\hskip-3.4pt N}
\newcommand{\zR}{I\hskip-3.4pt R}
\newcommand{\zw}{\omega}
\newcommand{\zW}{\Omega}
\newcommand{\zC}{{\mathbb C}}
\newcommand{\zD}{{\Delta}}
\newcommand{\zG}{{\Gamma}}
\newcommand{\veps}{\varepsilon}
\newcommand{\OM}{({\bf ***\ldots***})}
\newcommand{\EM}{({\bf $\leftarrow$***})}
\newcommand{\BM}{({\bf ***$\rightarrow$})}
\newcommand{\noi}{\noindent}

\newcommand {\bdm} {\begin{displaymath}}
\newcommand {\edm} {\end{displaymath}}
\newcommand {\ba}  {\begin{array}}
\newcommand {\ea}  {\end{array}}
\newcommand {\mapx} {\Phi_{\epsilon}}
\newcommand {\Hx}    {{\mathcal H}_{\hat x}}
\newcommand {\muje} {\mu_j(\eps)}
\newcommand {\ugns}[1] {{\bf u}^{gns}(#1)}
\newcommand {\ugnsi}[1] {{\bf u}^{gns}_i(#1)}
\newcommand {\uns}[1] {{\bf u}^{ns}(#1)}
\newcommand {\parno} {\par\noindent}
\newcommand {\para} [1]{ \par\noindent {\bf #1} \par\noindent}
\newcommand {\gam} [1]{\gamma_{{\vec k}_{#1}}}
\newcommand {\norgq} [1] {| \gamma_{{\vec k}_{#1}} |^2}

\newcommand{\td}{T_r}
\newcommand{\ts}{T_\ell}
\newcommand{\xt}{X_t}
\newcommand{\ft}{f_t}
\newcommand{\Ft}{{\cal F}_t}
\newcommand{\vt}{{\cal V}_t}
\newcommand{\kt}{{\cal K}_t}
\newcommand{\rot}{\rho_t}
\newcommand{\cL}{{\tt L}}
\newcommand{\cT}{{\cal T}^{(N)}}
\newcommand{\kT}{{\cal T}}
\newcommand{\cTa}{{\cal T}^\ast}
\newcommand{\se}{{(s)}}
\newcommand{\zP}{\mbox{P}}
\newcommand{\zM}{\mbox {M}}
\newcommand{\pp}{p^+}
\newcommand{\cmin}{\underline{\cal C}}
\newcommand{\cmed}{{\cal C}_{av}}
\newcommand{\cmax}{\bar{\cal C}}



\title{Asymmetric fluctuation-relaxation paths in FPU models}

\author{C. Giberti}
\affiliation{Dipartimento di Scienze e Metodi dell'Ingegneria,
Universit\`a di Modena e Reggio Emilia, Reggio Emilia, Italy}
\email{giberti@unimore.it}
\author{L. Rondoni}
\affiliation{Dipartimento di Matematica, Politecnico di Torino, Torino,
Italy}
\author{C. Vernia}
\affiliation{Dipartimento di Matematica,
Universit\`a di Modena e Reggio Emilia, Modena, Italy}


\pacs{05.40.-a, 45.50.-j, 02.50.Ey, 05.70.Ln, 05.45.-a}
\keywords{Deterministic vs. stochastic dynamics, large deviations,
irreversible thermodynamics}

\begin{abstract}
A recent theory by Bertini, De Sole, Gabrielli, Jona-Lasinio and Landim
predicts a temporal asymmetry in the fluctuation-relaxation paths of
certain observables of nonequilibrium systems in local thermodynamic
equilibrium. We find temporal asymmetries in the fluctuation-relaxation
paths of a form of local heat flow, in the nonequilibrium FPU-$\zb$ model
of Lepri, Livi and Politi.
\end{abstract}

\maketitle



\vskip 25pt


\section{Introduction}
In Refs.\cite{bsgj01,BDSJLcurrent}, Bertini, De Sole, Gabrielli, Jona-Lasinio and
Landim proposed a generalization of Onsager-Machlup's theory \cite{om53} to the
large fluctuations around nonequilibrium steady states. This theory predicts
temporal asymmetries between fluctuation and relaxation paths of certain
observables, and includes the results of Derrida, Lebowitz and Speer \cite{DLS},
and of Bodineau and Derrida \cite{BD} as special cases. In particular, the theory
of \cite{bsgj01,BDSJLcurrent} concerns stochastic lattice gases,
which admit the hydrodynamic description
\be
\partial_{t}\varrho = \nabla \cdot
\left[\frac{1}{2} D\left(\varrho \right)\nabla \varrho - \chi(\varrho) \nabla E
\right] \equiv \mathcal{D}\left(\varrho
\right) ~, \quad \quad \varrho = \varrho(u,t) ~,
\label{hydro}
\ee
where $\varrho$ is the vector of macroscopic observables, $u$ and $t$ are the
macroscopic space and time variables, $D$ is the Onsager diffusion matrix, $\chi$
is the linear response conductivity, and $\nabla E$ is the external driving
force. For such models, Refs.\cite{bsgj01,BDSJLcurrent} prove that the
spontaneous fluctuations out of a steady state most likely follow an {\em
adjoint} hydrodynamic equation:
\be
\partial_{t}\varrho =\mathcal{D}^{*}\left(\varrho \right) ~, \quad \mbox{ with }
\quad \mathcal{D}^*(\varrho) = \mathcal{D}(\varrho) - 2 \mathcal{A} ~,
\label{adjointH}
\ee
where $\mathcal{D}$ has been decomposed as
\be
\mathcal{D}(\varrho) = \frac{1}{2} \nabla \cdot \left( \chi(\varrho) \nabla
\frac{\delta \mathcal{S}}{\delta \varrho} \right) + \mathcal{A} ~,
\label{decompose}
\ee
and $\mathcal{A}$ is a vector field orthogonal to the thermodynamic force, i.e.\
to the functional derivative of the entropy with respect to the state
$\delta S/\delta \varrho$. The fact that $\mathcal{A}$ is orthogonal to the
thermodynamic force implies that it does not contribute to the entropy production:
$\mathcal{A}$ is the non-dissipative part of the dynamics.

Because the theory of \cite{bsgj01,BDSJLcurrent} is developed for
stochastic processes, the question arises whether its predictions can
be observed in the dynamics of time reversal invariant, deterministic,
particle systems, such as those of nonequilibrium molecular dynamics
(NEMD) \cite{EM}. Indeed, the stochastic description of a given system
is often thought to be a reduced representation of the microscopic
deterministic dynamics of its particles. One difficulty in addressing
this question is that not all reductions of a microscopic dynamics to
a mesoscopic (stochastic) one are equivalent, and it is not always
easy to identify the stochastic counterparts of the observables of
deterministic systems.

A nonequilbrium steady state requires at least two reservoirs at different
thermodynamic states, in contact with the system of interest, and NEMD
provides techniques to avoid the simulations of the reservoirs. However,
present day NEMD simulations cannot be performed with sufficiently many
particles, and for sufficiently long times to achieve the mesoscopic level
of description. Therefore, in Ref.\cite{GR04} the simplest of particle systems,
the nonequilibrium Lorentz gas, was investigated, but no temporal asymmetry
was found in the fluctuation-relaxation paths (FRPs) of its current. Here, we
consider the FRPs of the local heat flux of a more realistic model: the
nonequilibrium FPU-$\zb$ model of Lepri, Livi and Politi \cite{LLP1,LLP2}, and we
observe that they are asymmetric in time. The model consists of the usual
FPU-$\zb$ chain with $N$ anharmonic oscillators of equal masses $m$, located at
$x_j$, $j=1,..., N$, and with two Nos\'e-Hoover ``thermostats'' at different
temperatures acting on the first and the last oscillator. The interaction
potential and the internal energy are thus given by
\be
V(q)=\frac{q^2}{2} + \beta\frac{q^4}{4} ~; \quad \mbox{ and } \quad
H=\sum_{j=0}^N \frac 1 2 m \dot{q}_j^2+V(q_{j+1}-q_{j}) ~,
\label{potential}
\ee
where $q_j=x_j-ja$ is the displacement of the $j$-th oscillator from its
equilibrium position, and $a$ is the equilibrium distance between two nearest
neighbours. The equations of motion take the form
\bea
&& \hskip -25pt
\ddot{q}_1 = F(q_1 - q_0) - F(q_2-q_1) - \zeta_\ell \dot{q}_1 ~, \quad
\ddot{q}_N = F(q_N - q_{N-1}) - F(q_{N+1}-q_N) - \zeta_r \dot{q}_N \label{eqmot1} \\
&&\hskip -25pt
\ddot{q}_j = F(q_j - q_{j-1}) - F(q_{j+1}-q_j) ~, \quad
\mbox{for } j = 2,..., N-1 \label{eqsmot}
\eea
where $F(q) = - V'(q)$, and the terms $\zeta_\ell, \zeta_r$ obey
\be
\dot{\zeta}_r = \frac{1}{\theta_r^2} \left( \frac{\dot{q}_1^2}{T_r} -1 \right)
~, \quad \quad
\dot{\zeta}_\ell = \frac{1}{\theta_\ell^2} \left( \frac{\dot{q}_N^2}{\ts} -1
\right) ~,
\label{NHthermo}
\ee
with the boundary conditions $q_0 = q_{N+1} = 0$. We take $\zb=0.1$ and
$m=\theta_r=\theta_\ell=1$, as in \cite{LLP1}. Equations (\ref{eqmot1}) and
(\ref{NHthermo}) constitute the Nos\'e-Hoover thermostats at ``temperatures''
$T_r$ and $\ts$, with response times $\theta_r$ and $\theta_\ell$. This
nonequilbrium FPU-$\zb$ model is dissipative \footnote{The time average of
the divergence of the equations of motion,
$-(\langle \zeta_r \rangle + \langle \zeta_\ell \rangle)$, is negative.},but
time reversal invariant\footnote{There is an involution
$i: {\cal M} \to {\cal M}$, $i(q,p,\zeta) = (q,-p,-\zeta)$, $p=m\dot{q}$,
on the phase space ${\cal M}$, which anticommutes with the time evolution:
$iS^t = S^{-t} i$, where $S^t : {\cal M} \to {\cal M}$ is the time evolution.}.
Note that in our context $(T_r - T_\ell)$ is just a parameter expressing the
distance from the equilibrium state, and not a real temperature gradient.

Let $L$ be a number of contiguous oscillators in a sublattice
${\cL}=\{j_1,\ldots, j_L\}$ of the chain, whose center coincides with the center of the chain,
and let $\eta=L/N$ be the fraction of the chain occupied by $\cL$. The
instantaneous local heat flux ${\cal F}$ is defined by \cite{LLP2}:
\be
{\cal F}=\frac 1 L \sum_{j=j_1}^{j_L}\frac1
2(x_{j+1}-x_{j})(\dot{q}_{j+1}+\dot{q}_j) F(q_{j+1}-q_{j})+\dot{q}_j h_j,
\label{Heatflux}
\ee
where
\bdm
h_j=\frac 1 2 m \dot{q}_j^2+\frac 1 2 \left [ V(q_{j+1}-q_j)+V(q_j-q_{j-1})
\right ],
\edm
and $F(q)=-V^\prime(q)$. There is another definition of local heat flux,
which could be considered:
\be
f=\frac 1 L \sum_{j=j_1}^{j_L}\frac1 2 a F(q_j-q_{j-1})({\dot q}_j +
{\dot q}_{j-1})+\frac{1}{2L}a \left({\dot q}_{j_1}F(q_{j_1+1}-q_{j_1}) +
{\dot q}_{j_L}F (q_{j_L+1}-q_{j_L}) \right) ~,
\label{heatPD}
\ee
The first term of $f$ is the average over $\cL$ of
the heat flux in the limit of small (compared to the lattice spacing $a$)
oscillations as in \cite{LLP2}, while the second term gives the flux through
the boundary of $\cL$. The quantities ${\cal F}$ and $f$ are supposed to be
equivalent in the
limit of small oscillations and long chains. Therefore, for the sake of brevity,
in this paper we do not study the properties of the FRPs of $f$, which will be
considered in a future paper \cite{GRVn}, along with the FRPs of other
observables.

Because of the large difference between the mesoscopic level of description, and
the microscopic level simulated by us, it is not obvious that our FRPs are
directly related to those of \cite{bsgj01,BDSJLcurrent}; if they are not, their
temporal asymmetry is a feature of their dynamics which, to the best of our
knowledge, has not been studied before in microscopic deterministic dynamics.
However, the fact that our results do not change qualitatively with the growth
of $N$ suggests a relation between our results and the asymmetries of the large
deviations of Refs.\cite{bsgj01,BDSJLcurrent}, and indicates how the theory of
\cite{bsgj01,BDSJLcurrent} may extend to the microscopic level of description.

\section{Deterministic fluctuation-relaxation paths}
Given the observable ${\cal F}: {\cal M} \to \zR$, the identification of a FRP
around its steady state (expected) value $\zE({\cal F})$ requires some care.
Indeed, the time series of ${\cal F}$ looks very noisy, and takes the value
$\zE({\cal F})$ only at discrete instants of time. Similarly, the fluctuation
values selected to be observed are also achieved only at discrete instants of
time. Averaging over many particles, and over many microscopic times, stabilizes
the signal, but makes negligible the frequency of the FRPs, to the point that it
becomes impossible to assess their statistical properties.

By analogy with the notation of \cite{bsgj01}, we denote by ${\cal F}_t$ the
quantity ${\cal F}(S^t \zG)$, if $\zG$ is the initial point of the trajectory in
${\cal M}$, and $S^t : {\cal M} \to {\cal M}$ is the time evolution operator. We
want to identify the most likely FRP, according to the SNS
probability distribution, starting at ${\cal F}_i=\zE({\cal F})$, reaching the
fluctuation value ${\cal F}_T={\cal T}({\cal F})$, and returning to
${\cal F}_f=\zE({\cal F})$ after a certain time. To asses the dependence of the
phenomenon on $N$, we consider two kinds of fluctuation values:
\be
\cTa_s({\cal F})=\zE({\cal F})+s\,\sigma({\cal F}),\quad \mbox{ and } ~~
\cT_s({\cal F})=\zE({\cal F})+\frac{1}{s}
\sigma({\cal F})\sqrt{N}; \quad s=1,2,3...
\label{thersholds}
\ee
with $\sigma({\cal F})$ the standard deviation of ${\cal F}$ (which depends on
$N$), and we introduce the following definition of FRP:

\vskip 5pt
\noi
{\bf Definition 1. }{\it Let ${\cal F}$ take the value ${\cal T}({\cal F})$ at
time $\hat{t}$, with $[d {\cal F}_t / d t]_{\hat{t}} > 0$. Let $\bar{t}$ be the
smallest time after $\hat{t}$ such that ${\cal F}_{\bar{t}}={\cal T}({\cal F})$.
Choose a time $\zt_0 > 0$, representing the expected duration of the fluctuation
path. The union $\{ {\cal F}_{\hat{t}-\tau},\ \tau\in [0,\tau_0] \} \cup \{
{\cal F}_{\bar{t}+\tau},\ \tau\in [0,\tau_0] \}$ of the trajectory segments of
duration $\zt_0$ preceding $\hat{t}$, and following $\bar{t}$, is called an
{\em FRP}.
}

\vskip 5pt \noi
Numerically, the time $\hat{t}$ can be identified testing the condition
\be
({\cal F}_{(t+h)}-{\cal T})({\cal F}_t-{\cal T})<0,\quad
\mbox{where $h = $ time step, ~~ $\cal T = $ fluctuation value}.
\label{Ftime}
\ee
Whenever a FRP is obtained, we shift it in order to have $\hat{t}=0$.
Given $n$ FRPs, we collect them
in the set $\{(\tau, {\cal F}_\tau^{(s)}),\ s=1,2\ldots, n\}$, and build
a histogram by partitioning the rectangle
$[-\tau_0,\tau_0]\times [\min_{\tau,s} {\cal F}_\tau^\se,\
\max_{\tau,s} {\cal F}_\tau^\se]$ with small rectangular bins, and
evaluating the frequency of visitation of each bin. The number of temporal
intervals in which $[-\zt_0,\zt_0]$ is subdivided is denoted by $b_\tau$, while
the number of bins of $[\min_{\tau,s} {\cal F}_\tau^\se,\ \max_{\tau,s}
{\cal F}_\tau^\se]$ is denoted by $b_{\cal F}$. The area of each
rectangular bin $\Delta_{p,q}$, with $p=1,\ldots,b_\tau,\ q=1,\ldots,b_{\cal F}$,
is $\Delta=2\zt_0 [\max_{\tau,s} {\cal F}_\tau^\se-\min_{\tau,s}
{\cal F}_\tau^\se]/(b_\tau b_{\cal F})$, and the corresponding histogram is a
two-dimensional surface, whose ``crest'', defined by
\be
{\cal C}(p)=\left((2\tau_0+1)n\Delta\right )^{-1}\max_{q=1,\ldots,b_{\cal F}}\sharp
\left\{(\tau,{\cal F}_\tau^\se) \in \Delta_{p,q},\ s=1,2\ldots, n\right\},\quad
p=1,\ldots,b_\tau ,
\label{crest}
\ee
with $\sharp \{ . \}$ the number of FRPs in $\{ . \}$, represents the
most likely FRP in the sample. To assess the asymmetries of the FRPs,
and of the corresponding crests, we introduce the following definition:

\vskip 5pt
\noi
{\bf Definition 2. }{\it Let $\kT$ be the fluctuation value chosen for ${\cal F}$.
If $\check{{\cal F}}(\hat{t},\bar{t},\zt_0)=\{ {\cal F}_{\hat{t}-\tau},\
\tau\in [0,\tau_0] \} \cup \{ {\cal F}_{\bar{t}+\tau},\ \tau\in [0,\tau_0] \}$ is
an FRP; its {\em asymmetry coefficient} is given by
\be
\za(\check{{\cal F}}(\hat{t},\bar{t},\zt_0))=\frac{1}{\tau_0(\kT-\zE({\cal F}))}
\sum_{0 \le \tau \le \tau_0} \left( {\cal F}_{\bar{t}+\tau} - {\cal F}_{\hat{t}-\tau}
\right).
\label{FRasymmIII}
\ee

\noi
Let ${\cal C}: \{1,\ldots b_\tau\} \to \zR$ be the crest obtained form the
given sample of FRPs; the {\em asymmetry coefficient} of ${\cal C}$ is given by
\be
\alpha_c({\cal C})=\frac{2}{b_\tau(\kT-\zE({\cal F}))}\left(
\sum_{p=\frac{b_\tau}{2}+1}^{b_\tau} {\cal C}(p) -
\sum_{p=1}^{\frac{b_\tau}{2}} {\cal C}(p) \right ).
\label{crestasymm}
\ee

\noi
The asymmetry coefficient $\za$ of a single FRP is defined in the same way,
replacing the values ${\cal C}(p)$ with the corresponding values of the FRP.
A FRP or a crest is symmetric if its asymmetry coefficient vanishes.
}

\noi
Observe that the asymmetry coefficients can take positive as well as negative
values, because the differences which define them are not taken in absolute value.
This is motivated by the fact that random oscillations of a fluctuation path
around its symmetric relaxation path should result in a vanishing
asymmetry coefficient.

\section{Numerical results}
We deal with an observable which is odd under the time reversal operation.
Unfortunately, the crests obtained from our simulations are strongly noisy,
as evidenced by Fig.1 in one of the cases considered, and their asymmetry
coefficient $\za_c$ is not particularly meaningful. Therefore, we study the
statistics of the asymmetry coefficient $\za$ of the single FRP, considered as
a random variable which takes different values on the different paths. Our
results are summarized in Tables I and II, in which the fraction
of oscillators over which the averages are computed to the total number of
oscillators are fixed, but the other parameters defining the model vary.
\begin{figure}
\centering
               \includegraphics[width=6.5cm,height=6.5cm]{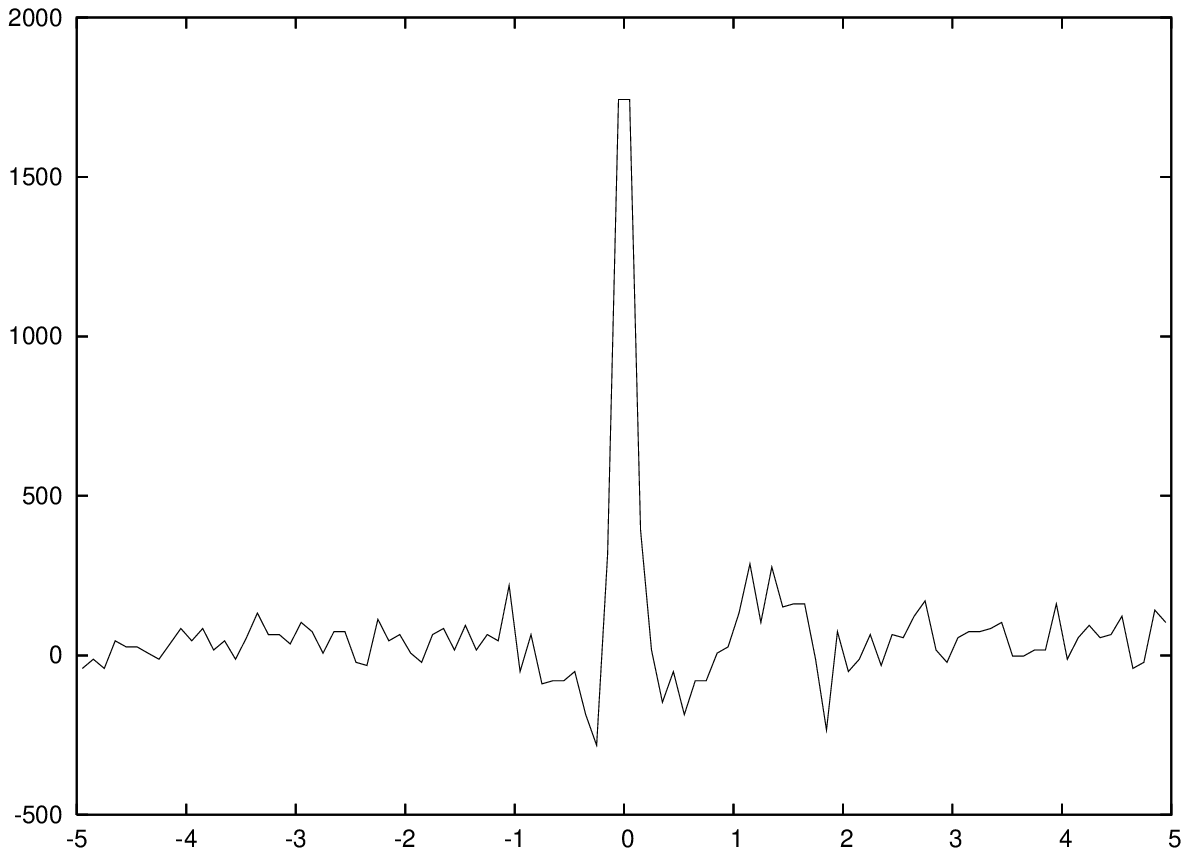}
               \caption{\small Crest of the observable ${\cal F}$, with $N=100,\eta=0.16,
               \ts=200,\td=20$,
               $\zE(f)\simeq 24.741$, and $\cTa_3\simeq 1738$. The asymmetry coefficient of the crest,
               computed according to (\ref{FRasymmIII}), is $\za_c({\cal C})=0.0142$.}
\label{20infnity}
\end{figure}
We conclude that the most likely FRP, which is the object of the theory of
\cite{bsgj01,BDSJLcurrent}, is hard to assess directly, because the crest
obtained for each different case is highly noisy and not sharp: indeed, the
supports and the variances of the marginal probability distributions, for
fixed $\zt$ in the flux space, are quite wide (cf.\ Tables I and II).
\begin{table}
\centerline{\begin{tabular}{|c|c|c|c|c|c|c|}\hline $N$ & $\zE({\cal
F})$& ${\sigma({\cal F})}$ & $P(\za>0)$ ; $\cTa_3$ &
$\zE(\za)$;$\cTa$ & $P(\za > 0)$ ; $\cT_4$ & $\zE(\za)$ ; $\cT_4$ \\
\hline 25& 49.2262& 1449.5949 &0.7001 &0.0254 & - & -  \\
\hline 50& 37.3129& 777.1178&0.6316  &0.0336 & - & -\\
\hline 100& 26.4482 & 569.0022&0.6354  &0.0391 &0.6200 &0.0447 \\
\hline 150& 21.3395 & 405.9141&0.6121  &0.0492 &0.6103 &0.0483 \\
\hline 200&18.2231 & 355.5294&0.6142  &0.0478 &0.6074 &0.0410 \\
\hline 250&15.7638 & 309.9113&0.6122  &0.0481 &0.6075 &0.0388 \\
\hline 300&14.7470 & 295.3437&0.5987  &0.0410 &0.5951 &0.0384 \\
\hline 350&12.9422 & 271.1457& 0.5865  &0.0350 &0.5845 &0.0386 \\
\hline 400& 12.1784& 249.9643&0.5960  &0.0347 &0.5732 &0.0319 \\
\hline \end{tabular}} \caption{\small Mean and standard deviation of
${\cal F}$, probability of positive asymmetry with different
fluctuation values, and mean of its single FRPs, as functions of
$N$. Here, $\eta=0.16$, $\ts=200,\ \td=20$, and $\tau_0=15$.}
\label{table6}
\end{table}
\begin{table}
\centerline{\begin{tabular}{|c|c|c|c|c|}\hline $N$
& $\zE({\cal F})$& ${\sigma({\cal F})}$ & $P(\za>0)$ &
$\zE(\alpha)$\\
\hline 250 & 25.8539& 564.2185&0.5840&0.0305\\
\hline 300 &23.8107 &558.0274 &0.6082&0.0334\\
\hline 350 &24.0250 &571.7431&0.6219&0.0484\\
\hline 400 &20.3576 &554.6692 &0.6480&0.0430\\
\hline $350^{(*)}$&40.6265&1117.7722&0.6297&0.0474\\
\hline $400^{(*)}$ &41.4094 &1099.1702 &0.6131&0.0473\\
\hline \end{tabular}} \caption{\small Same as Table I, for $\cT_4$
only, with $\ts=300$. The cases with a (*) have
$\ts=500$.}\label{table6.1}
\end{table}

\noi
On the other hand, all the data Tables I and II
indicate that the FR paths of ${\cal
F}$ have a positive asymmetry, like those of $f$ \cite{GRVn}.
A positive asymmetry of the FRPs is
testified not only by the fact that the mean asymmetry is a positive
number, but especially by the fact that the probability of being postive
is well above $1/2$. Furthermore, the dependence on the various
parameters defining the model is rather weak.

Therefore, we have found evidence of a new phenomenon in the nonequilibrium
FPU-$\zb$ model of Lepri, Livi and Politi, which shows how irreversible
phenomena can be obtained out of (dissipative) reversible dynamics.
Furthermore, the weak dependence of our results on $N$ suggests that
the asymmetries may survive in the large $N$ limit, with growing fluctuation
values, although
at present it is not possible to test directly very large $N$'s and $\kT$'s.
If the asymmetries really survive, our results constitute a first (partial)
verification of the predictions of Refs.\cite{bsgj01,BDSJLcurrent} (originally
obtained for nonequilibrium stochastic systems), in the context of time reversal
invariant, deterministic dissipative particle systems.

\section*{Acknowledgements}
The authors are grateful to G.\ Jona-Lasinio, D.\ Isbister and R.\ Livi for
enlightening discussions.

\end{document}